\documentclass[10pt,letterpaper,twocolumn,australian,superscriptaddress,twocolumns,10pt,citeautoscript,nofootinbib]{revtex4-1}

\usepackage[T1]{fontenc}
\usepackage{color}
\usepackage{babel}
\usepackage{booktabs}
\usepackage{amsmath}
\usepackage{amssymb}
\usepackage{graphicx}
\usepackage{wasysym}
\usepackage{esint}
\usepackage[unicode=true,pdfusetitle,
 bookmarks=true,bookmarksnumbered=false,bookmarksopen=false,
 breaklinks=false,pdfborder={0 0 1},backref=false,colorlinks=false]
 {hyperref}

\makeatletter

\pdfpageheight\paperheight
\pdfpagewidth\paperwidth

\providecommand{\tabularnewline}{\\}

\PassOptionsToPackage{caption=false}{subfig}

\newcommand{\ket}[1]{\left| #1 \right>}
\newcommand{\bra}[1]{\left< #1 \right|}

\newcommand{\expect}[1]{\left< #1 \right>}

\let\oldsqrt\sqrt
\def\sqrt{\mathpalette\DHLhksqrt}
\def\DHLhksqrt#1#2{%
\setbox0=\hbox{$#1\oldsqrt{#2\,}$}\dimen0=\ht0
\advance\dimen0-0.2\ht0
\setbox2=\hbox{\vrule height\ht0 depth -\dimen0}%
{\box0\lower0.4pt\box2}}

\newcommand{\appropto}{\mathrel{\vcenter{   \offinterlineskip\halign{\hfil$##$\cr     \propto\cr\noalign{\kern2pt}\sim\cr\noalign{\kern-2pt}}}}}

\usepackage{siunitx}

\@ifundefined{showcaptionsetup}{}{%
 \PassOptionsToPackage{caption=false}{subfig}}
\usepackage{subfig}
\makeatother

\begin{document}

\title{Characterisation of an exchange-based two-qubit gate for resonant
exchange qubits}

\author{Matthew P. Wardrop}

\author{Andrew C. Doherty}

\affiliation{Centre for Engineered Quantum Systems, School of Physics, The University
of Sydney, Sydney, NSW 2006, Australia}

\date{\today}
\begin{abstract}
Resonant exchange qubits are a promising addition to the family of
experimentally implemented encodings of single qubits using semiconductor
quantum dots.  We have shown previously that it ought to be straightforward
to perform a CPHASE gate between two resonant exchange qubits with a single exchange pulse. This approach uses energy
gaps to suppress leakage rather than conventional pulse sequences. In this
paper we present analysis and simulations of our proposed two-qubit gate
subject to charge and Overhauser field noise at levels observed in current experiments. Our
main result is that we expect implementations of our two-qubit gate to achieve high fidelities,
with errors at the percent level and gate times comparable to single-qubit
operations. As such, exchange-coupled resonant
exchange qubits remain an attractive approach for quantum computing.
\end{abstract}
\maketitle

\section{Introduction}

The seminal work of Loss and Divincenzo\cite{Loss1998} introduced the notion of
using individual electrons trapped in gate-defined quantum dots to encode
quantum information, an idea which has since burgeoned into a family of
promising architectures for quantum computing
\cite{Petta2005,Koppens2006,Nowack2007,Pioro-Ladriere2008,Barthel2009,Foletti2009,Shulman2012,Laird2010,Gaudreau2011,Medford2013,Kloeffel2013}.

An early theoretical realisation was that a single qubit encoded in three
electron spins could be universally controlled using exchange couplings alone
\cite{DiVincenzo2000a}, which removes any requirement for individually
addressable electron spin resonance or magnetic field gradients. Crucially,
since exchange couplings in semiconductor experiments are controllable using
gate voltages, this allows all qubit operations to be performed electronically;
an attractive feature in experimental implementations. Single-qubit operations
for the so-called ``exchange-only'' qubit have been experimentally demonstrated
\cite{Medford2013,Medford2013a}. Pulse sequences are known for single-qubit
gates that simultaneously correct leakage errors and other sources of
noise~\cite{Hickman2013}. Proposed two-qubit gates for the exchange-only qubit
either involve capacitive coupling \cite{Pal2015,Pal2014} or exchange-coupling
\cite{DiVincenzo2000a,Kawano2005,Fong2011,Setiawan2014}. Since exchange coupling
is usually much larger than capacitive coupling, exchange gates are usually
faster, but come at the cost of requiring complicated pulse sequences in order
to echo away the unwanted spin-flip transitions that occur as a side-effect and
cause leakage errors. The first such pulse sequence \cite{DiVincenzo2000a} that
effected a CNOT required 19 exchange pulses in 13 timesteps, and was found using
a numerical search. Since then, robust numerical searches have found improved
pulse sequences that are robust against more sources of decoherence and/or
reduce the number of gate operations \cite{Kawano2005,Fong2011,Setiawan2014}.

A recent alternative to the ``exchange-only'' qubit is the ``resonant-exchange
qubit'' \cite{Taylor2013,Medford2013}, which encodes qubits in the
interaction picture with respect to significant exchange coupling
between the three dots. Universal single qubit operations are effected
using rf gate pulses to the electrodes controlling the exchange couplings.
This qubit has been shown both theoretically \cite{Taylor2013} and
experimentally \cite{Medford2013} to have several improved properties,
including first-order insensitivity to charge fluctuations and reduced
leakage error due to nuclear field fluctuations \cite{Medford2013,Fei2015}.
In addition, Taylor et al. \cite{Taylor2013} have shown that you
can perform two-qubit gates between these qubits using charge dipole
interactions, while we have suggested an alternative of using simple
exchange pulses between nearby qubits \cite{Doherty2013}.

In this earlier work \cite{Doherty2013}, we showed that a two-qubit
CPHASE gate can be implemented using a single exchange pulse between
the constituent quantum dots of neighbouring qubits (shown schematically
in figure \ref{fig:Physical-configurations}). Rapid high-fidelity
gate operation is in principle made possible by energetically suppressing
the spin-flip transitions that lead to leakage. This method of effecting
a two-qubit gate contrasted with the more conventional approach of
long and complicated pulse sequences, and is similar to our earlier
proposal for singlet-triplet qubits \cite{Wardrop2014}.

It is the purpose of this work to extend our previous results by considering
higher-order analysis, adiabatic pulse profiles, the effect of noise,
and the performance of our gate in physically motivated simulations
that include noise. In section \ref{sec:Resonant-Exchange-Qubits}
we briefly review single-qubit resonant-exchange qubit operations,
in section \ref{sec:Our-Two-Qubit} we review our two-qubit gates,
in section \ref{sec:Gate-Characterisation} we formally characterise
our two-qubit gate in the butterfly geometry, in section \ref{sec:Alternative-Geometries}
we briefly consider other geometries, and in section \ref{sec:Conclusions}
we conclude.

In this work we set $\hbar=1$, meaning that energies are interchangeable
with angular frequencies.

\section{Resonant Exchange Qubits\label{sec:Resonant-Exchange-Qubits}}

\begin{figure}
\includegraphics[width=1\columnwidth]{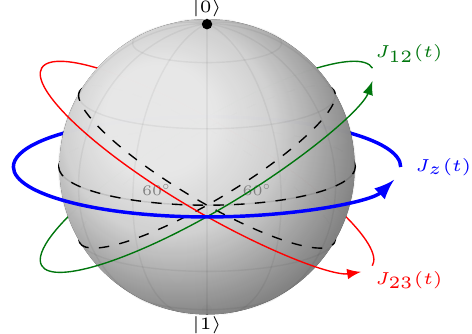}

\protect\caption{(colour online) A Bloch sphere schematic depicting the effect of intra-qubit
exchange couplings on qubit states. Under $J_{12}$ and $J_{23}$
couplings, states rotate about the same axes as those of the rotations
indicated in green and red respectively. If both $J_{12}$ and $J_{23}$
are equal to $J_{z}$, states rotate about the $z$ axis as indicated
by the emboldened blue rotation. Pulses involving different $J_{12}$
and $J_{23}$ couplings allow arbitrary rotations, and hence provide
universal single qubit control. As these intra-qubit couplings are
controllable using gate voltages, this allows for universal electronic
control of qubits.\label{fig:single-qubit-control}}

\end{figure}

A resonant exchange qubit is a triple-dot system operating deep in the (1,1,1)
charge state (each dot almost surely confines a single electron). A large
magnetic field is applied along the $z$-axis, Zeeman splitting the $2^{3}=8$
spin states according to the z-projection of their total spins: $\Delta
E=-m_{z}B_{\perp}$, $B_{\perp}$ is the effective Zeeman splitting of an electron
subject to the global magnetic field. Note that we have absorbed the g-factor
(up to its sign) into our definition of $B_{\perp}$. The logical states of the
qubit are
$\ket{0}=\left(\ket{\uparrow\uparrow\downarrow}+\ket{\downarrow\uparrow\uparrow}-2\ket{\uparrow\downarrow\uparrow}\right)/\sqrt{6}$
and
$\ket{1}=\left(\ket{\uparrow\uparrow\downarrow}-\ket{\downarrow\uparrow\uparrow}\right)/\sqrt{2}$,
which both have total spin $S=1/2$ with z-projection of $m_{z}=1/2$. The
remainder of the eight-dimensional Hilbert space describes non-logical states,
the spanning eigenstates of which are completed for the energy eigenbasis in
table \ref{tab:qubit-states}. These qubits are operated with large intra-qubit
exchange couplings $J_{12}$ and $J_{23}$, with oscillatory modulations around
$J_{12}=J_{23}=J_{z}$ providing single qubit control
\cite{Taylor2013,Medford2013}. This is depicted and described in more detail in
figure \ref{fig:single-qubit-control}.
 Intra-qubit exchange couplings are in turn controlled by detuning
the voltages defining the quantum dots (which we parametrise as $\varepsilon$).
This allows for complete electronic control of qubits, which potentially
simplifies experimental implementation. In several GaAs singlet-triplet
qubit experiments \cite{Petta2005,Foletti2009,Laird2010,Dial2013},
an exponential ansatz $J(\varepsilon)=J_{0}\exp(\varepsilon/\varepsilon_{0})$
has been found to be a good phenomenological fit to experimental data
over a wide range of interesting values of $\varepsilon$, and so
we adopt it in this work.
\begin{table}
\begin{centering}
\begin{tabular}{ccccc}
\toprule
Label & State & $S$ & $m_{z}$ & Energy\tabularnewline
\midrule
$\ket{Q_{3/2}}$ & $\ket{\uparrow\uparrow\uparrow}$ & 3/2 & 3/2 & $-3B_{\perp}/2$\tabularnewline
$\ket{0}$ & $\ket{\uparrow\uparrow\downarrow}+\ket{\downarrow\uparrow\uparrow}-2\ket{\uparrow\downarrow\uparrow}$ & 1/2 & 1/2 & $-B_{\perp}/2-3J_{z}/2$\tabularnewline
$\ket{1}$ & $\ket{\uparrow\uparrow\downarrow}-\ket{\downarrow\uparrow\uparrow}$ & 1/2 & 1/2 & $-B_{\perp}/2-J_{z}/2$\tabularnewline
$\ket{Q}$ & $\ket{\uparrow\uparrow\downarrow}+\ket{\uparrow\downarrow\uparrow}+\ket{\downarrow\uparrow\uparrow}$ & 3/2 & 1/2 & $-B_{\perp}/2$\tabularnewline
$\ket{0_{-}}$ & $\ket{\downarrow\downarrow\uparrow}+\ket{\uparrow\downarrow\downarrow}-2\ket{\downarrow\uparrow\downarrow}$ & 1/2 & -1/2 & $B_{\perp}/2-3J_{z}/2$\tabularnewline
$\ket{1_{-}}$ & $\ket{\downarrow\downarrow\uparrow}-\ket{\uparrow\downarrow\downarrow}$ & 1/2 & -1/2 & $B_{\perp}/2-J_{z}/2$\tabularnewline
$\ket{Q_{-}}$ & $\ket{\downarrow\downarrow\uparrow}+\ket{\downarrow\uparrow\downarrow}+\ket{\uparrow\downarrow\downarrow}$ & 3/2 & -1/2 & $B_{\perp}/2$\tabularnewline
$\ket{Q_{-3/2}}$ & $\ket{\downarrow\downarrow\downarrow}$ & 3/2 & -3/2 & $3B_{\perp}/2$\tabularnewline
\bottomrule
\end{tabular}
\par\end{centering}

\protect\caption{Energy eigenstates of a single resonant exchange qubit, sorted by
$m_{z}$. $S$ is the total angular momentum quantum number of the
three electron spins and $m_{z}$ is the z-component of the total
angular momentum. $J_{z}=J_{12}=J_{23}$ is the energy level splitting
due to exchange coupling, and $B_{\perp}$ is the splitting of an
electron subject to the large transverse global field. Each eigenstate
with a negative subscript has their constituent spins flipped relative
to the corresponding unsubscripted state.\label{tab:qubit-states}}
\end{table}

\section{Exchange-Coupled Two-Qubit Gate\label{sec:Our-Two-Qubit}}

\begin{figure}
\includegraphics[width=1\columnwidth]{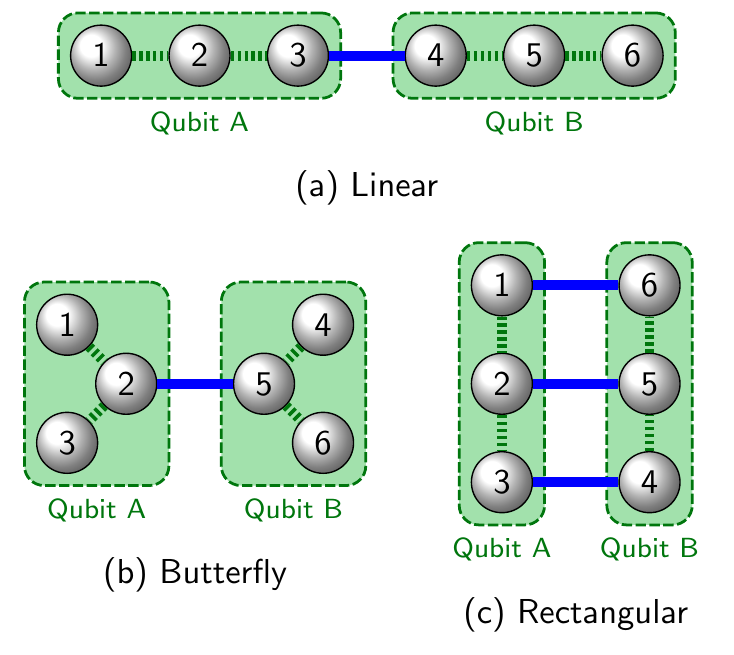}\subfloat{\label{subfig:linear}}\subfloat{\label{subfig:butterfly}}\subfloat{\label{subfig:rectangular}}

\protect\caption{(colour online) The physical arrangements of quantum dots (or geometries)
considered in this work. Dashed green lines indicate large intra-qubit
exchange couplings, with solid blue lines indicating the weaker exchange
coupling between qubits that can be used to effect two-qubit gates.
We will focus mainly on the (a) butterfly and (b) linear and geometries,
with the (c) rectangular geometry being another alternative. \label{fig:Physical-configurations}}
\end{figure}

Here we provide a brief review of the two-qubit gate between resonant
exchange qubits described in our prior letter \cite{Doherty2013}.
Consider two resonant-exchange qubits ($A$ and $B$) in a large transverse
magnetic field, coupled in several different ways as depicted in figure
\ref{fig:Physical-configurations}
. We label the intra-qubit couplings of the qubits $J_{z}^{A}$ and
$J_{z}^{B}$, and assume that the couplings within each qubit are
equal (i.e. $J_{12}^{A,B}=J_{23}^{A,B}=J_{z}^{A,B}$). The logical
states of the two qubit system, $\ket{0,0}$, $\ket{0,1}$, $\ket{1,0}$
and $\ket{1,1}$, are all in the $m_{z}=1$ subspace, along with eleven
other states. Since we will only be considering dynamics which conserve
z-projection of spin $m_{z}$ and are working in a large global magnetic
field, this reduces the dimension of the Hilbert space that can interact
with the logical subspace to a maximum of $15$; though in some cases
(such as for the butterfly geometry) symmetry constrains this subspace
further. For an explicit representation of all $15$ energy levels
in the energy eigenbasis, please refer to the supplementary material.
The most important observation to make about the non-logical $m_{z}=1$
subspace is that they all have energies different from those on the
logical subspace by at least $\min(J_{z}^{A},J_{z}^{B},B_{\perp})$.
This guarantees that leakage transitions will be unfavourable provided
that all additional energy-level splittings remain less than $\sim J_{z}$,
which leads to times of order at least $\sim1/J_{z}$. Since single
qubit operations already run slow compared to these timescales , this
is not a restriction in practice.

Using the fact that leakage processes are suppressed when $J_{c}\ll J_{z}$,
we performed first order degenerate perturbation theory around $J_{c}=0$
and wrote down its effect on the logical subspace; the so-called effective
Hamiltonian\cite{Cohen-Tannoudji2008} on the logical subspace. The
zeroth order terms arising from perturbation theory describe the uncoupled
resonant exchange qubits. Writing the Pauli-Z logical operators on
qubit A and B as $\sigma_{z}^{A}$ and $\sigma_{z}^{B}$ respectively,
the zeroth order effective Hamiltonian is:
\begin{equation}
H_{0}=-(J_{z}^{A}+J_{z}^{B})-\frac{1}{2}J_{z}^{A}\sigma_{z}^{A}-\frac{1}{2}J_{z}^{B}\sigma_{z}^{B}.\label{eq:effective-intra-hamiltonian}
\end{equation}
The first order terms describe the leading order effect of the inter-qubit
coupling $J_{c}$. The effect on the logical subspace is described
by:
\begin{eqnarray}
H_{c} & = & \delta J_{c}+\frac{1}{2}\delta J_{z}(\sigma_{z}^{A}+\sigma_{z}^{B})\nonumber \\
 &  & +J_{zz}\sigma_{z}^{A}\sigma_{z}^{B}+J_{\perp}(\sigma_{x}^{A}\sigma_{x}^{A}+\sigma_{y}^{B}\sigma_{y}^{B}),\label{eq:effective-coupling-hamitonian}
\end{eqnarray}
with $J_{zz}$, $\delta J_{z}$ and $J_{\perp}$ all being geometry
dependent, as specified in table \ref{tab:Qubit-coupling-parameters}.
\begin{table}
\begin{centering}
\begin{tabular}{cccc}
\toprule
Geometry & $\delta J_{z}/J_{c}$ & $J_{zz}/J_{c}$ & $J_{\perp}/J_{c}$\tabularnewline
\midrule
Linear & $1/36$ & $1/36$ & $-1/24$\tabularnewline
Butterfly & $-1/18$ & $1/9$ & $0$\tabularnewline
Rectangular & 0 & $1/6$ & $-1/12$\tabularnewline
\bottomrule
\end{tabular}
\par\end{centering}

\protect\caption{Qubit coupling parameters arising from lowest order perturbation theory
in each of the three geometries of figure \ref{fig:Physical-configurations}.
Calculations assume that $J^{A}_{z}\simeq J^{B}_{z}$, and that all non-zero
exchange couplings $J_{ij}$ are equal to $J_{c}$. When $\left|J^{B}_{z}-J^{A}_{z}\right|\gg J_{c}$,
the degeneracy of the logical $\ket{10}$ and $\ket{01}$ states is
broken, and we find that $J_{\perp}\rightarrow0$ for all geometries.
All other entries in the table are unaffected.\label{tab:Qubit-coupling-parameters}}

\end{table}
 This perturbative analysis is repeated in greater detail in the supplementary
material.

The structure of these effective Hamiltonians admit a straightforward
two-qubit CPHASE gate using a single DC exchange pulse. Simple AC
coupling pulses may also be interesting \cite{Doherty2013}, but we
leave this to future work. Consider first two qubits coupled according
to the butterfly geometry of figure \ref{subfig:butterfly}. Since
$J_{\perp}=0$ in this geometry, the only two-qubit component of the
gate's operation at first order is $J_{zz}$, which will implement
a CPHASE gate after a time $\tau$ such that $\int_{0}^{\tau}J_{zz}dt=\pi/4$
(modulo single qubit unitaries). The other two geometries, linear
and rectangular in figures \ref{subfig:linear} and \ref{subfig:rectangular}
respectively, have non-zero $J_{\perp}$; and consequently will not
perform a CPHASE gate unless the additional contribution can be suppressed.
This can be achieved by detuning the intra-qubit exchange coupling
energies such that $\left|J_{z}^{B}-J_{z}^{A}\right|\gg J_{c}$; or
by adding a simple logical Z ($\sigma_{z}^{A}$ or $\sigma_{z}^{B}$)
echo pulse at $t=\tau/2$ to one of the qubits associated with each
exchange coupling (which anti-commutes with $\sigma_{x}^{A}\sigma_{x}^{B}$
and $\sigma_{y}^{A}\sigma_{y}^{B}$, and thus cancels out the effect
of $J_{\perp}$). We opt not to consider more sophisticated pulse
sequences that echo out higher order contributions to $J_{\perp}$
in order to maintain the simplicity of our gate.

\section{Gate Characterisation\label{sec:Gate-Characterisation}}

We now begin a more complete characterisation of the performance of
our two-qubit gate. The gate has two intrinsic sources of error (which
would be present even in an ideal implementation): timing inaccuracies
and leakage; and we will consider the two sources of extrinsic noise
anticipated to be most pertinent in experimental implementation: charge
and Overhauser noise.

In the following two subsections, we will show that intrinsic noise
can be effectively mitigated by correctly tuning gate times and by
adiabatic pulse sequences, resulting in high fidelity gate operations.
We then move on to consider how robust our gate is to the anticipated
sources of experimental noise. Simulations will be provided in each
section to demonstrate the anticipated performance of our gate. These
simulations involve monte-carlo averaging (over pseudo-static parameters)
of solutions to a lindblad master equation (encoding high frequency
noise). The performance measure used is ``entanglement fidelity'',
as described in a former work \cite{Wardrop2014}. ``Entanglement
fidelity'' is related to the more commonly used ``average fidelity''
of random benchmarking by:
\[
\bar{F}=\frac{dF_{e}+1}{d+1},
\]
where $\bar{F}$ is the average fidelity, $F_{e}$ is the entanglement
fidelity, and $d$ is the dimension of the quantum system \cite{Horodecki1999,Nielsen2002}
($d=4$ for our two-qubit system). Entanglement fidelity is used in
this work because it can be directly computed using a fixed input
state, which simplifies simulations. In this section, the simulations
are usually done for the butterfly geometry which has the greatest
symmetry and performance. In the next section (\ref{sec:Alternative-Geometries}),
we extend our analysis to the linear geometry, which should be easier
to fabricate for experiment.

\subsection{Timing Inaccuracies\label{sub:Estimation-of-Ideal-Gate-Time}}

Due to the complexity of the dynamics of the six quantum dot system,
there is no closed analytic form for the ideal gate time. As a result,
one needs to be careful how the gate time is estimated; over- or under-estimating
the ideal gate time will result in a corresponding over or under accrual
of two qubit phase, and thus reduced gate fidelities.

The ideal gate time is the time $\tau$ for which a noiseless exchange
pulse should be turned on between the two triple-quantum-dot systems
in order to perform a two-qubit CPHASE gate on the encoded qubits.
Recall from section \ref{sec:Our-Two-Qubit} that $\tau$ is implicitly
defined by $\int_{0}^{\tau}J_{zz}dt=\pi/4$, with $J_{zz}$ being
geometry (and potentially time) dependent, as shown in table \ref{tab:Qubit-coupling-parameters}.

Using the butterfly configuration as an example, first order perturbation
theory predicts that $J_{zz}(J_{c})=J_{c}/9$ and thus implies that
$\tau=9\pi\expect{J_{c}}_{t}/4$; where $\expect{J_{c}}_{t}$ is the
time average of $J_{c}$ during the pulse. Note that the linearity
of the first order approximation for $J_{zz}(J_{c})$ allows one to
compute gate time $\tau$ in a manner agnostic to the details of the
pulse shape, requiring knowledge only of the average value of $J_{c}$
during the pulse. This property is lost beyond first order, as $J_{zz}(J_{c})$
has corrections at higher order that become significant in all geometries
for physically relevant values of $J_{c}$ and $J_{z}$, meaning that
$\tau$ must be calibrated anew for each pulse shape.

There is no closed analytic form for $\tau$, and since $\tau$ would
in any case have to be calibrated in-situ in any experimental implementation
using one of several optimisation techniques \cite{Kelly2014,Egger2014,Ferrie2014,Granade2014},
we refer the reader to the supplementary material for a description
of how we numerically optimise $\tau$ in our simulations. Henceforth,
we assume $\tau$ has been estimated perfectly, and note that in many
of our simulations $\tau$ differs significantly from its first order
estimates.

\subsection{Leakage\label{sub:Leakage}}

Leakage is a measure of how much a state initially with support only
on the logical subspace shifts support onto the non-logical subspace
during a logical operation, and results in reduced entanglement fidelities
(to first non-trivial order, $\mathcal{F}=1-\mathcal{L}$). For an
arbitrary state $\rho$, we quantify this using $\mathcal{L}=\mathrm{Tr}(P\rho P)$,
where $P=1-\sum_{l=00,01,10,11}\ket{l}\bra{l}$ is the projector off
the logical subspace. Leakage occurs via energetically forbidden excitations
that are suppressed by the energy gap between the logical states and
a leakage state, or when logical states are subjected to pulses with
frequencies corresponding to the energy gap.

\begin{figure}
\includegraphics[width=0.8\columnwidth]{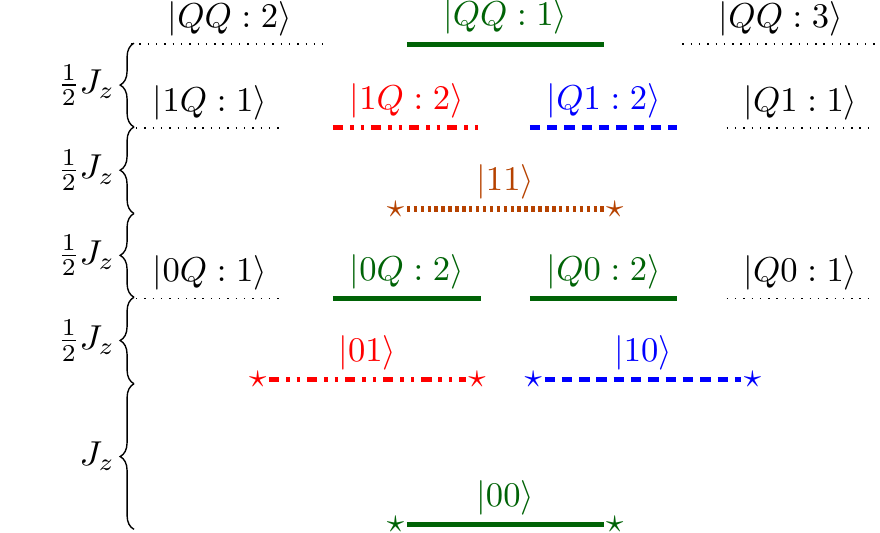}

\protect\caption{(colour online) The energy level spectrum of the $m_{z}=1$ subspace
of a two resonant exchange qubit system. Energy levels are each labelled,
with starred levels indicating logical eigenstates. The eigenstates
corresponding to each label are explicitly written out in the supplementary
material. Under butterfly coupling $J_{c}$, the logical states couple
to like coloured/dashed states leading to a minimal energy gap of
$\sim3J_{z}/2$. Note that the like coloured/dashed levels form distinct
subspaces. Under linear coupling, all coloured/thick energy levels
are coupled (including logical states), leading to a minimal energy
gap of $\sim J_{z}/2$. \label{fig:Energy-Levels}}
\end{figure}

In the two-triple-quantum-dot system, logical states are isolated
from leakage states by an energy gap $\Delta E$ proportional to $J_{z}$.
When coupled using $J_{c}$, this energy gap monotonically reduces.
Consequently, as the ratio $J_{c}/J_{z}$ increases, the likelihood
of leakage also increases. The energy level spectrum, with an indication
as to which non-logical states the logical states can couple, is shown
in figure \ref{fig:Energy-Levels}. Note that the symmetries of the
butterfly geometry cause $J_{c}$ to couple logical states to disjoint
subspaces and give rise to an effective energy gap of $\Delta E=3J_{z}/2$,
which is three times larger than in the linear system where $\Delta E=J_{z}/2$
due to $J_{c}$ coupling all of the logical states into the same subspace.
This leads to substantially improved performance in the butterfly
configuration for any given $J_{c}/J_{z}$.

\begin{figure}

\includegraphics[width=1\columnwidth]{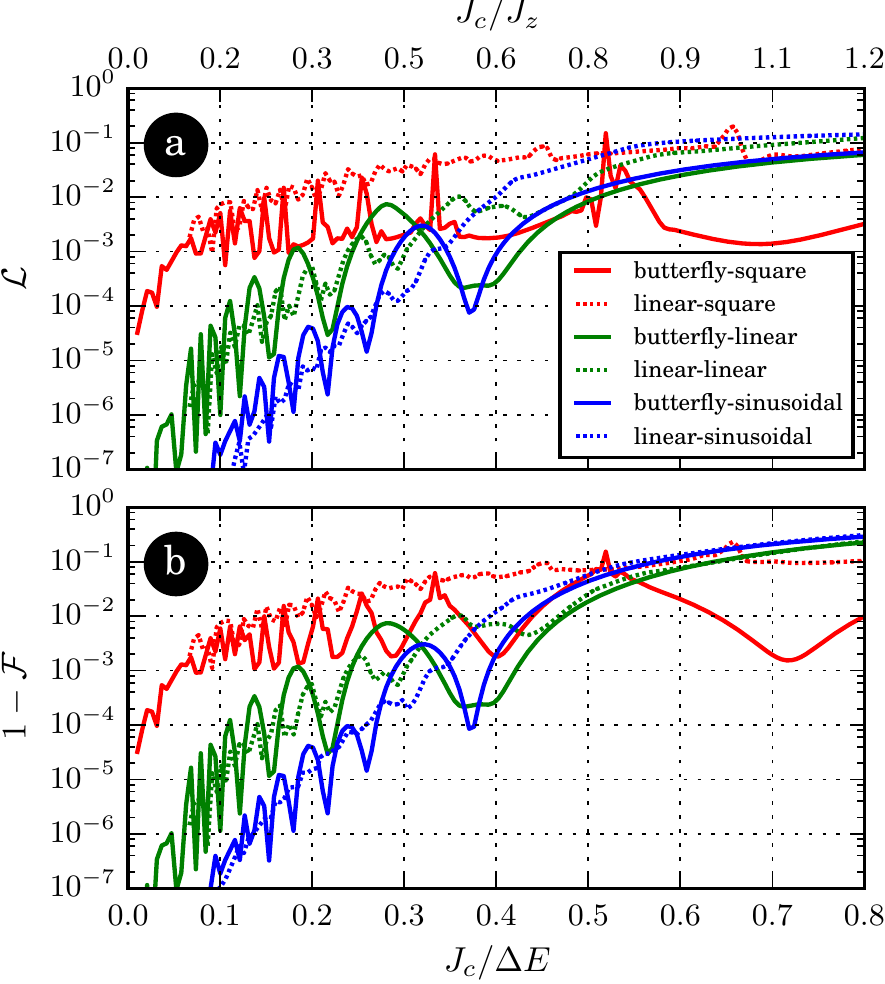}

\protect\caption{(colour online) (a) Leakage $\mathcal{L}$ and (b) infidelity $1-\mathcal{F}$
(bottom) at the end of a single two-qubit gate operation for several
different adiabatic profiles in both the butterfly and linear geometries.
The x-axis is shared between the plots, and is over $J_{c}/\Delta E$:
the ratio of inter-qubit coupling and the geometry-dependent minimum
energy gap. The upper x-axis provides a conversion from $J_{c}/\Delta E$
to $J_{c}/J_{z}$ for the butterfly geometry (an expression in terms
of the controllable parameters of the model). Colours indicate the
adiabatic pulse profile used, while solid (dashed) lines denote that
the butterfly (linear) geometry is being considered. Crucially, these
plots demonstrate that use of adiabatic pulses can improve suppression
of leakage by several orders of magnitude, provided that $J_{c}$
is small compared to the energy gap; and that this leads to a corresponding
increase in gate fidelities. \label{fig:leakage-limited-performance}}
\end{figure}

Implementing our two-qubit gate requires $J_{c}$ to be active only
for a fixed duration $\tau$, meaning that rapid or broadband changes
in $J_{c}$ when it is turned on and off can lead to excitations from
the logical subspace. Choosing pulse shapes with discontinuities only
at high differential orders can therefore further suppress leakage
by several orders of magnitude, and hence improve fidelities, as shown
in figure \ref{fig:leakage-limited-performance}.
 For more intuition regarding adiabatic pulses and leakage refer to
our earlier work on adiabatic pulses for singlet-triplet qubits \cite{Wardrop2014}.

In the absence of noise, the entanglement fidelity $\mathcal{F}\approx1-\mathcal{L}$
is limited only by leakage, which is in turn determined by the choice
of adiabatic profile and the ratio $J_{c}/J_{z}$. This provides an
upper bound on the performance of implementations of our gate. In
this work, we will use a narrow-band sinusoidal adiabatic pulse described
by:
\[
\tilde{J_{c}}=J_{\mathrm{c}}\left(1-\cos\left(\frac{2\pi t}{\tau}\right)\right),
\]
which has its first discontinuity at second order when $t=0$ and
$\tau$. Note that $\tilde{J_{c}}$ is chosen such that $J_{c}$ is
the average value of the pulse. This choice allows for fast single-pulse
gates with fidelities in excess of $0.9999$ for physically reasonable
parameters, as shown in figure \ref{fig:leakage-limited-performance},
and hence this gate may prove to be useful for fault-tolerant computation
using semiconductor quantum dots. The remainder of this section will
be devoted to determining how robust this performance is to anticipated
sources of experimental noise.

\subsection{Charge Noise}

\begin{table}
\begin{tabular}{cc}
\toprule
Parameter & Value\tabularnewline
\midrule
Exchange couplings: & \tabularnewline
$J_{z}=J^{A}_{z}=J^{B}_{z}$ & \SI{1.65}{\micro\eV} (\SI{\sim0.4}{\GHz})\tabularnewline
Exponential ansatz: & \tabularnewline
$J_{0}$ & \SI{82.7}{\micro\eV} (\SI{\sim20}{GHz})\tabularnewline
$\varepsilon_{D}$ & \SI{0.35}{\mV}\tabularnewline
Charge noise: & \tabularnewline
$\sigma_{\varepsilon}$ & \SI{15.8}{\micro\V}\tabularnewline
$D$ & \SI{0.244}{\micro\V^2\ns}\tabularnewline
Overhauser noise: & \tabularnewline
$\sigma_{B}$ & \SI{2}{mT}\tabularnewline
\bottomrule
\end{tabular}

\protect\caption{Model parameters. In the experiments of Medford et al. \cite{Medford2013},
the intra-qubit couplings $J_{z}$ had measured values of roughly
\SI{0.8}{\micro eV} (\SI{0.2}{\GHz}) to \SI{4}{\micro eV} (\SI{1}{GHz}),
of which we've chosen a conservative value. The parameters for the
exponential ansatz $J(\varepsilon)=J_{0}\exp(\varepsilon/\varepsilon_{0})$
were chosen to roughly match the experimental results of Dial and
collaborators \cite{Dial2013}. The noise parameters $\sigma_{\varepsilon}$
and $D$ were calibrated by respectively matching somewhat typical
values of $T_{2}^{*}=\SI{25}{ns}$ and $T_{2}=\SI{20}{\micro s}$
from experiment \cite{Medford2013,Medford2013a}. The standard deviation
of the Overhauser field $\sigma_{B}$ was chosen to be consistent
with \cite{Koppens2006,Koppens2007,Nowack2007,Reilly2008a}.\label{tab:Noise-parameters}}
\end{table}

One of the most significant sources of experimental noise affecting
semiconductor quantum dot qubits is charge noise\textcolor{red}{{} }\cite{Dial2013,Medford2013}.
Charge noise is the effect of fluctuations in electric potential on
the gates defining the quantum dots, for which there are many causes
including environmental rf radiation and Johnson noise \cite{Beaudoin2014}.
Charge noise on the electrodes exhibits itself in our model as fluctuations
in the electrode detunings $\varepsilon$.

\begin{figure*}
\includegraphics[width=0.8\textwidth]{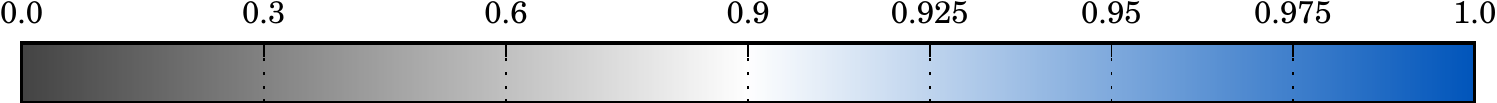}

\subfloat{\protect\includegraphics[width=1\columnwidth]{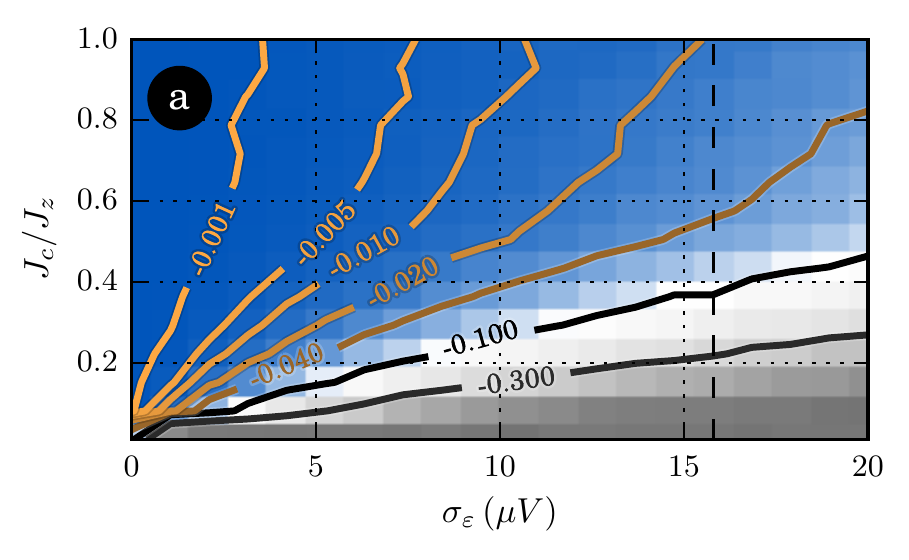}\label{fig:dc-charge-noise}}\subfloat{\protect\includegraphics[width=1\columnwidth]{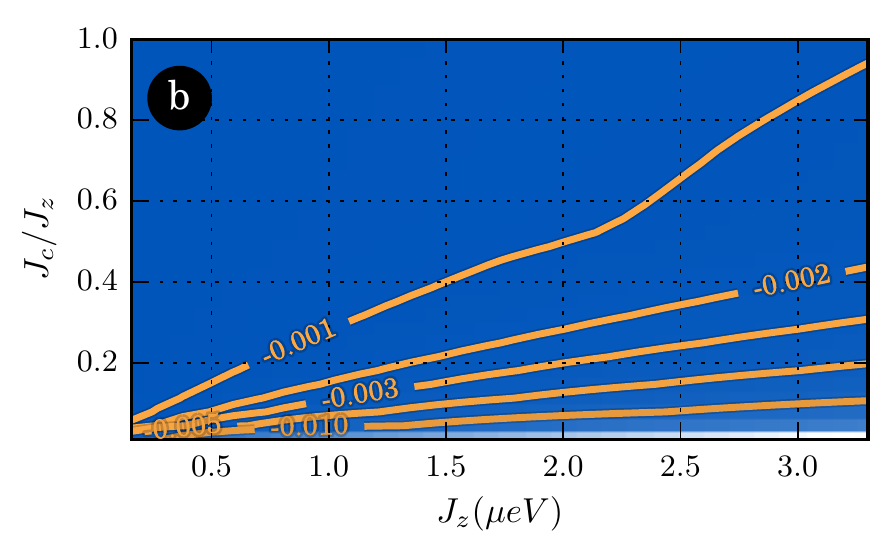}\label{fig:hf-charge-noise}}\protect\caption{(colour online) The contribution to fidelity ($\mathcal{F}_{\mathrm{noisy}}-\mathcal{F}_{\mathrm{noiseless}}$)
from (a) pseudo-static (DC) and (b) white (HF) charge noise on each
of the five exchange couplings active during a single gate operation
using the butterfly geometry. The colourmap used is divergent at a
fidelity of 0.9, as shown above; with blue colours indicating performance
in excess of 0.9, and grey colors indicating performance below 0.9.
Since DC noise is essentially dependent only on the the ratio $J_{c}/J_{z}$,
we plot in (a) the infidelity contribution as a function of $J_{c}/J_{z}$
and $\sigma_{\varepsilon}$. The value of $\sigma_{\varepsilon}$
from table \ref{tab:Noise-parameters} is indicated by a dashed line
at $15.8\,\mu V$. In (b) we plot the fidelity contribution from white
charge noise for $D=0.244\,\mu V^{2}ns$ as in table \ref{tab:Noise-parameters}.
The qualitative behaviour of both plots is predicted by equation \ref{eq:fidelity-approx-noisy}.
\label{fig:Effect-of-charge}}
\end{figure*}

In this work, we describe charge noise on $\varepsilon$ using a two-parameter
phenomenological model that approximates the noise in gate voltages
by static and white noise perturbations around the desired value.
The pseduo-static (DC) and white noise (HF) perturbations are respectively
parameterised by the standard deviation of the pseudo-static charge
offset $\sigma_{\varepsilon}$ and the spectral density of charge
fluctuations $D$; which can be respectively calibrated to experimental
$T_{2}^{*}$ and $T_{2}$ characteristic times. This is the same model
described in our earlier work on singlet-triplet qubits \cite{Wardrop2014},
which seems to reasonably describe the results of experiments \cite{Dial2013},
even though the precise mechanisms that cause this behaviour are not
perfectly understood \cite{Beaudoin2014}. In experiment, there are
likely to be additional high-frequency $T_{1}$ processes biased toward
relaxation, which we have chosen not to include in this model as it
would require adding parameters to our model that have not been sufficiently
well empirically constrained. Moreover, it has been found in experiment
that characteristic $T_{1}$ times are in excess of $\sim\SI{40}{\micro s}$
for $J_{z}\lesssim\SI{1.5}{\micro eV}$ \cite{Medford2013}, which
is long compared to $T_{2}$ times of $T_{2}\sim\SI{20}{\micro s}$
\cite{Medford2013}. Even though $T_{1}$ times are found to decrease
with larger $J_{z}$ \cite{Medford2013}, and so at some point will
become comparable to the dominant sources of noise, the performance
of our gate does not appear in any case to be limited by high frequency
noise in the parameter space of interest to us (see figure \ref{fig:hf-charge-noise}).
We therefore do not expect that the fidelity of our gate operations
will be strongly affected by $T_{1}$ relaxations.

During the operation of our gate, there are multiple exchange couplings active
at once (five in the case of the butterfly and linear geometries). We assume
that charge noise is independent on each coupling, and use as mentioned earlier
an exponential ansatz for each coupling:
$J_{ij}(\varepsilon_{ij})=J_{0}\exp(\varepsilon_{ij}/\varepsilon_{0})$, where
$i$ and $j$ indicate the pair of dots being coupled. We also assume that the
size and spacing of each pair of dots is the same, allowing us to use the same
$J_{0}$ and $\varepsilon_{0}$ for each coupling. $J_{0}$ and $\varepsilon_{0}$
are calibrated to match such that $J(\varepsilon)$ is consistent with the data
from Dial et al. \cite{Dial2013}. In the small noise limit in which we are
interested, first order analysis of the effect of perturbations to the
intra-qubit couplings $J_{12}$ and $J_{23}$ allow one to derive that
$\sigma_{\varepsilon}=\sqrt{2}\hbar\varepsilon_{0}/J/T_{2}^{*}$ and
$D=2\hbar^{2}\varepsilon_{0}^{2}/J^{2}/T_{2}$; where $J$ is the exchange
coupling inferred from the experiment. In this work, we fit these parameters to
experimental single resonant-exchange qubit $T_{2}$ times of \SI{\sim
20}{\micro\s} and $T_{2}^{*}$ times of \SI{\sim 25}{\ns}
\cite{Medford2013,Medford2013a}. The resulting parameters are shown in table
\ref{tab:Noise-parameters}.

The simplicity of the noise model and exponential ansatz allows us
to construct a qualitative model from perturbation theory of the effects
of this noise model on the entanglement fidelity at the end of a single
two-qubit gate operation:

\begin{eqnarray}
\mathcal{F} & \simeq & 1-\mathcal{L}-\frac{k^{2}}{16\varepsilon_{0}^{2}}\left(1+\sqrt{2}J_{z}^{2}/J_{c}^{2}\right)\left(\sigma_{\varepsilon}^{2}+DJ_{c}/k\right),\label{eq:fidelity-approx-noisy}
\end{eqnarray}
where $\mathcal{L}$ is the ultimate leakage, $J_{c}$ and $J_{z}$
are the inter- and intra-qubit exchange couplings respectively, $\varepsilon_{0}$
is parameter of the exponential ansatz and $k$ is a geometry dependent
constant term.

This qualitative model provides several key insights. Firstly, we learn from the
$\left(1+\sqrt{2}J_{z}^{2}/J_{c}^{2}\right)$ factor that the fidelity dimunition
due to noise becomes more significant for longer gate times ($J_{c}/J_{z}$
small), and so one is encouraged to run the gate as quickly as possible. The
$\sqrt{2}J_{z}^{2}/J_{c}^{2}$ term arises from DC noise on intra-qubit
couplings, and proves to be the dominant contribution to infidelity. The
$\left(\sigma_{\varepsilon}^{2}+DJ_{c}/k\right)$ factor suggests that DC noise
(parameterised by $\sigma_{\varepsilon}$) leads to an appoximately uniform
dimunition of fidelity for any given $J_{c}/J_{z}$, whereas high frequency noise
(parameterised by $D$) becomes relatively more significant as $J_{c}$ increases;
and that their effect is additive. Putting this all together we predict that
gate performance decreases when $J_{c}/J_{z}$ is too large (where leakage errors
dominate) or too small (where low frequency charge noise dominates). For the
experimentally relevant parameters considered in this work, we will not enter a
regime where $J_c$ is sufficiently large that high frequency charge noise
dominates. These predictions are corroborated in simulations of the butterfly
geometry including high frequency and DC charge noise, as shown in figure
\ref{fig:Effect-of-charge}. With the noise parameters chosen to correspond to
current experiments, we note that the gate's performance in the presence of
charge noise appears to be limited by intra-qubit DC noise. This pseudo-static
noise can in principle be echoed out by a single echo pulse. We will consider
the effect of such simple echo pulses later when discussing the linear geometry.
As well as echo pulses, we expect that technical developments will reduce the
level of low frequency charge noise in the future.

\subsection{Overhauser Noise}

Perhaps the most widely used substrate in semiconductor quantum dot
experiments is the GaAs/AlGaAs heterostructure. The nuclei in this
substrate have non-zero spin, which give rise to a net magnetisation
which slowly (compared to gate times) varies due to nuclear spin flip-flop
interactions \cite{Taylor2007}. This net polarisation is called an
Overhauser field. While there exist semiconductor substrates composed
of nuclei which do not have a spin (such as silicon and graphene),
GaAs/AlGaAs has remained a popular material due to the well-developed
fabrication techniques associated with it. In this section, we quantify
the effect of varying levels of Overhauser noise on the performance
of our gate.

The Overhauser field looks like an additional randomly oriented local
magnetic field at each dot. We assume that only a small fraction of
the $\sim10^{6}$ nuclei in the vicinity of each dot also contribute
significantly to the polarisation of an adjacent dot, and therefore
make the approximation that these local fields are uncorrelated. In
GaAs/AlGaAs structures, the Overhauser field has an RMS magnitude
of about \SIrange[range-units=single]{1}{3}{mT} \cite{Reilly2008a}.

We model the Overhauser field as a pseudo-static offset of the magnetic
field along the $z$-axis sampled from a normal distribution with
standard deviation $\sigma_{B}$. The neglected in-plane components
of the random field contribute at second order in perturbation theory,
and their effect is suppressed when the system is subject to a large
magnetic field along $z$ (as already posited in earlier sections),
and so can be safely ignored in this analysis. For a more complete
analysis, we refer the reader to Hung et al. \cite{Hung2014}.

The effect of this random Overhauser field is to couple the logical
subspace to all of the fifteen $m_{z}=1$ states shown in figure \ref{fig:Energy-Levels},
leading to both leakage and logical errors. The logical errors directly
caused by the Overhauser field are limited to single qubit errors,
as must be the case since the magnetic field perturbations are local
to each dot.

\begin{figure*}
\subfloat{\protect\includegraphics[width=1\columnwidth]{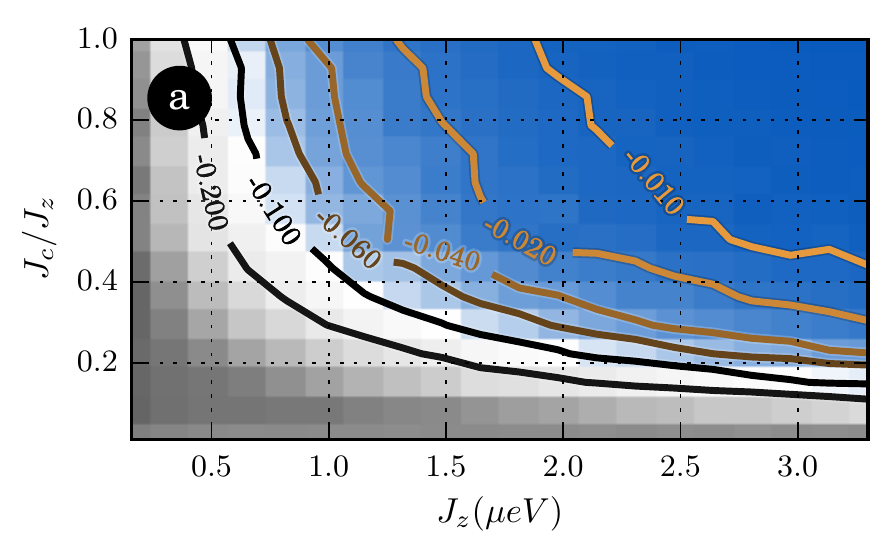}\label{fig:overhauser-Jc-vs-Jz}}\subfloat{\protect\includegraphics[width=1\columnwidth]{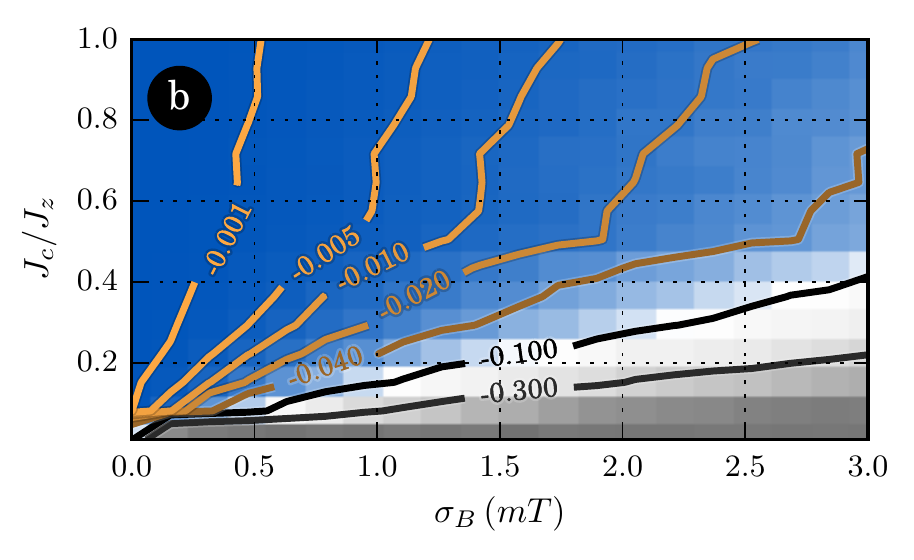}\label{fig:overhauser-Jc-vs-StdB}}\protect\caption{(colour online) The contribution to fidelity ($\mathcal{F}_{\mathrm{noisy}}-\mathcal{F}_{\mathrm{noiseless}}$)
from a pseudo-static Overhauser field with standard deviation $\sigma_{B}$.
In (a) the fidelity contribution is plotted as a function of $J_{c}/J_{z}$
and $J_{z}$ for $\sigma_{B}=2\,mT$, revealing that Overhauser noise
contributes to infidelity in a manner somewhat proportional to gate
time (contours of constant gate time are indicated by grey lines),
except at large $J_{c}/J_{z}$ corresponding to large leakage. In
(b) the fidelity contribution is plotted as a function of $J_{c}/J_{z}$
(with fixed $J_{z}=1.65\,\mu eV$) and $\sigma_{B}$. The radial contours
of (b) (in regions where leakage does not dominate) imply that infidelity
is proportional to $\sigma_{B}/J_{c}$ which is roughly $\propto\tau\sigma_{B}$,
thus corroborating that the infidelity grows roughly as gate time.
The colourmap used is the same as in figure \ref{fig:Effect-of-charge}.
\label{fig:Overhauser-Noise-Plots}}
\end{figure*}
 Simulations of entanglement fidelity for a range of different Overhauser
field magnitudes in an otherwise noiseless gate implementation are
shown in figure \ref{fig:Overhauser-Noise-Plots}. Our monte-carlo
simulations preferentially sample some magnetic field gradients in
order to increase the rate of convergence, as described in the supplementary
material. From \ref{fig:overhauser-Jc-vs-Jz}, we see that the effect
of the Overhauser field is approximately linear in gate time (breaking
down only once leakage becomes significant). From \ref{fig:overhauser-Jc-vs-StdB},
we see its effect is also approximately linear in the standard deviation
of the field.

It should be noted that various techniques exist for reducing the
magnitude of the Overhauser field in GaAs/AlGaAs, for example by nuclear
state preparation which has been shown to reduce the RMS by a factor
of $\sim70$ \cite{Reilly2008}; in which case the effects of the
Overhauser field can be largely neglected. Implementing these techniques
can be quite complicated, however, and so we assume conservatively
that the field will be unsuppressed .

\subsection{Cumulative Noise Model}

\begin{figure}
\includegraphics[width=1\columnwidth]{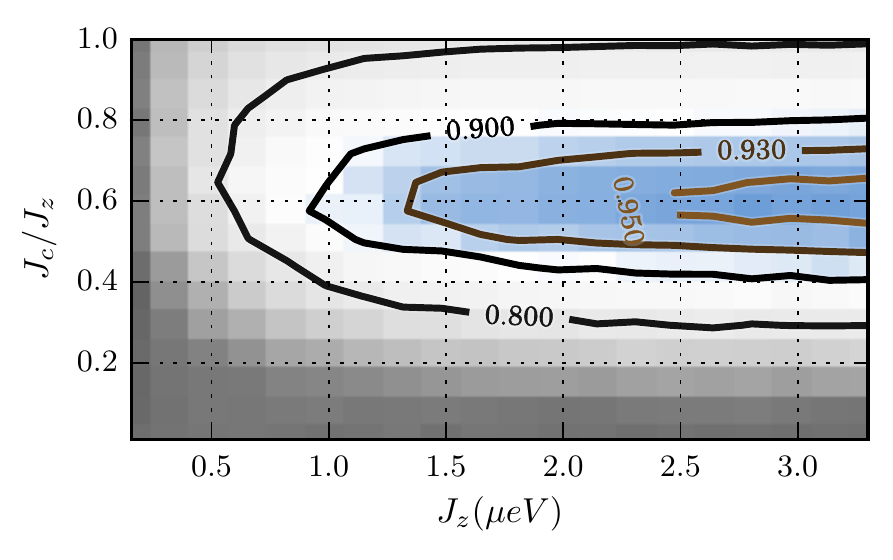}

\protect\caption{(colour online) Entanglement fidelity for the butterfly geometry subject
to charge and Overhauser noise, as described in the main text, with
noise levels calibrated to correspond to experiment (see table \ref{tab:Noise-parameters}).
Performance is found to be limited predominantly by leakage (from
above), intra-qubit DC charge noise (from below), and Overhauser noise
(from the left); leading to an optimal ratio of $J_{c}/J_{z}\sim0.6$
where fidelity reaches $\sim95\%$ for large enough $J_{z}$. The
colourmap used is the same as in figure \ref{fig:Effect-of-charge}. }
\end{figure}
As mentioned earlier, charge and Overhauser noise are the two most
significant sources of noise in semiconductor quantum dot experiments,
and in this section we simulate the performance of our gate in the
presence of both.

In principle the effects of Overhauser and charge noise can be more serious
than for either noise source alone. For example, we learn from perturbation
theory that at second order and above,
the Overhauser field cross-couples with charge noise in $J_{c}$,
allowing the Overhauser field to effect genuine two-qubit errors.
The larger $J_{c}/J_{z}$ and $\mathrm{\sigma_{B}}$ become, the greater
this cross-coupling and hence the non-linearity in the contribution
of charge noise and Overhauser noise to fidelity. Fortunately, this
effect is small in the parameter regime in which we are interested, and so we
refer the reader to the supplementary information for more details. This is
visible in the simulations discussed below, in which the effect of charge and
Overhauser noise on gate fidelities is (very nearly) additive.

A simulation of the performance of our two-qubit gate in the presence
of both charge and Overhauser noise is shown in figure \ref{fig:Overhauser-Noise-Plots}.
In this figure we see that the trade-off of avoiding leakage at large
$J_{c}/J_{z}$, while running the gate fast enough to avoid the accumulation
of DC charge noise and Overhauser noise, leads to an optimal value
for $J_{c}/J_{z}$ of approximately $0.6$ for large enough $J_{z}$.
Significantly, using just a single exchange pulse, we predict fidelities
of $\sim0.95$ with parameters currently accessible in experiment.

This optimal value of $J_c/J_z$ is surprisingly large compared to intuitions
garnered from prior arguments based on lowest order perturbation theory
~\cite{Doherty2013}, in which one must satisfy $J_c \ll J_z$. When this tighter
contraint is satisfied, implementations of this operation would actually take
longer than a more traditional approach involving multiple exchange
pulses~\cite{Setiawan2014}; and as seen in the results of our simulations, would
in any case have low fidelities. That high fidelities are achievable with a large 
optimal value of $J_c\simeq 0.6 J_z$ is one of the main results of our study, 
and is due to the tuning of gate times and use of adiabatic pulses discussed 
in sections \ref{sub:Estimation-of-Ideal-Gate-Time} and \ref{sub:Leakage} respectively.

There is still a large gap between the gate fidelities found in the presence of
these realistic noise sources and the upper limit imposed by leakage, as
discussed in section \ref{sub:Leakage}. This could be mitigated in several ways.
Firstly, one could increase our conservative choice of $J_{z}$, with larger
values of $J_z$ leading to higher fidelities; with the caveat that it has been
observed in experiment that qubit relaxation rates increase with $J_z$, likely
due to phonons. More experimental investigation would be required to determine
the optimal choice of $J_z$, but experimental evidence~\cite{Medford2013} suggests that the
parameters we have chosen are not too far from optimal for current devices.
Secondly, one could add a spin echo pulse at $\tau/2$, which will be discussed
in more detail for the linear geometry in the next section. For the butterfly
geometry, simulations involving an echo pulse boost fidelities to $\sim0.98$ at
the cost of slowing down the gate.  Thirdly, fidelities could be increased by
suppressing the Overhauser field using DNP, or using a material without nuclear
spin. We found that suppressing the Overhauser field entirely while also echoing
the low frequency charge noise allowed for fidelities exceeding $0.99$.

\section{Alternative Geometries\label{sec:Alternative-Geometries}}

While we have so far focussed mainly on the butterfly geometry due
to its simplifying symmetries and higher performance, it seems likely
that it will be much easier to fabricate a linear array of quantum
dots. We therefore extend our characterisation to include the anticipated
performance of a linear geometry, as shown in figure \ref{fig:Physical-configurations}.

In section \ref{sub:Leakage}, we noted that the linear exchange coupling
$J_{34}$ lacked the symmetry of the butterfly coupling, leading to an effective
energy gap three times smaller than the butterfly geometry. Figure
\ref{fig:leakage-limited-performance} demonstrated that when this is taken into
account in the absence of experimental noise, and leakage is plotted as a
function of $J_{c}/\Delta E$, the leakage of the butterfly and linear
arrangements is qualitatively similar. As detailed in section
\ref{sec:Our-Two-Qubit}, the $J_{\perp}$ term in equation
\ref{eq:effective-coupling-hamitonian} is non-zero for the linear geometry,
which makes it necessary to apply an echo pulse that anticommutes with the
$\sigma_{x}^{A}\sigma_{x}^{A}+\sigma_{y}^{B}\sigma_{y}^{B}$ contribution at
$t=0$ and at $t=\tau/2$ in order to perform a
CPHASE gate~\cite{Doherty2013}. One such choice (which we adopt in this work) is
a $pi$ rotation of the $B$ qubit about the z-axis, which effects a $\sigma_z^B$ operation.
This could be experimentally
implemented for the resonant exchange qubit by briefly shifting the value of
$J_{zB}$. Since $J_{c}$ changes the energy spectrum of the logical subspace, it
is either necessary to ensure that $J_{c}$ is turned off when the echo is
applied. For our choice of a sinusoidal pulse envelope amounts to halving the
time of the pulse and repeating it twice. Prior to each of these $J_c$ pulses
one performs the single qubit $\sigma_{zB}$ operation.

\begin{figure}
\includegraphics[width=1\columnwidth]{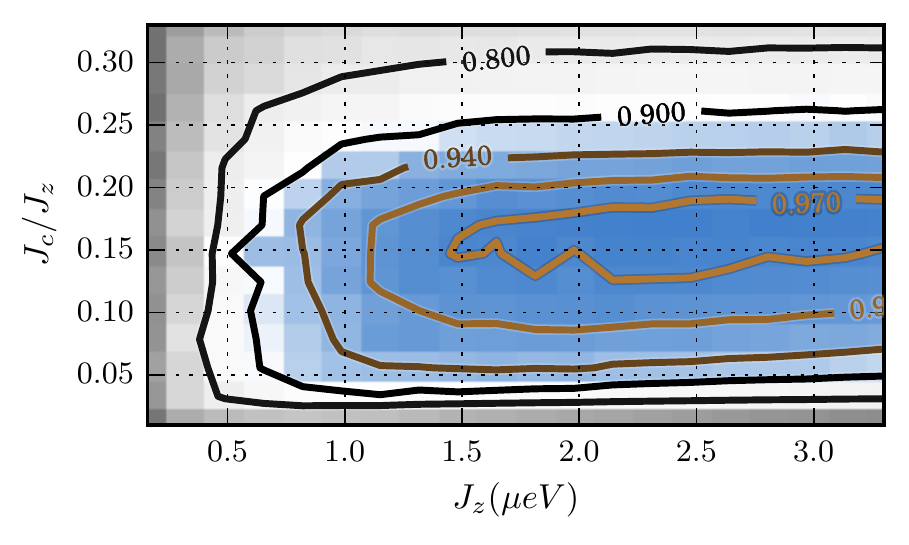}

\protect\caption{(colour online) Entanglement fidelity for the linear geometry subject
to charge and Overhauser noise, as described in the main text, with
noise levels calibrated to correspond to experiment (see table \ref{tab:Noise-parameters}).
A single $\sigma_{x}^{A}\sigma_{y}^{B}$ echo pulse is applied at
$\tau/2$ in order to echo out both intra-qubit psuedo-static charge
noise and non $\sigma_{z}^{A}\sigma_{z}^{B}$ two-qubit phase, which
restores maximum fidelities to $\sim97\%$. Fidelities are limited
predominantly by leakage (from above), Overhauser (from left and below),
and high frequency noise (from below). The colourmap used is the same
as in figure \ref{fig:Effect-of-charge}. \label{fig:linear-performance}}
\end{figure}

The smaller energy gap $\Delta E$ means that the gate must be run more slowly
than in the butterfly geometry, making the linear geometry more susceptible to
noise. The most significant contribution to infidelity
in figure \ref{fig:dc-charge-noise} was DC noise on the intra-qubit
couplings, which is geometry independent and diminishes fidelities
for $J_{c}/J_{z}\apprle0.5$. Since leakage becomes significant in
the linear geometry for $J_{c}/J_{z}\apprge0.3$, without suppressing
DC noise, high-fidelity operation is not possible without modifying the gate.
Here we demonstrate that suppression of DC noise is possible by modifying the
echo pulse used. Applying a pulse at $t=0$ and $t=\tau/2$ that anticommutes with
the $\sigma_{z}^{A}$ and $\sigma_{z}^{B}$ terms of equation
\ref{eq:effective-intra-hamiltonian} (such as $\sigma_{x}^{A}\sigma_{x}^{B}$)
will echo out the effect of DC noise on the intra-qubit couplings. This can be
merged with the $\sigma_{z}^{B}$ pulse discussed above. In this work we choose
to apply a $\sigma_x^A\sigma_x^B$ pulse, which becomes (up to irrelevant phase)
$\sigma_x^A\sigma_y^B$  when merged with $\sigma_z^B$. Note that this amounts to
an independent simultaneous $\pi$  rotation for each qubit. In resonant exchange
qubits this pulse can be achieved  by oscillatory $J$ pulses that are chosen to
be out of phase by $\pi/2$.  In the simulations shown in figure
\ref{fig:linear-performance} we demonstrate that these pulse sequences result in
improved fidelities of $\sim97\%$.

Similar to the butterfly geometry, gate performance is reduced for
large and small values of $J_{c}/J_{z}$ by leakage and (inter-qubit)
DC charge noise respectively; and for small values of $J_{z}$ and/or
$J_{c}/J_{z}$ by Overhauser noise.

In this analysis, we have assumed that the echo pulses (which are single qubit
operations) are instantaneous and performed with higher fidelity than two-qubit
operations. In actual fact, current experiments demonstrate single qubit gate
times several times longer \cite{Medford2013} than the duration of our two-qubit
gate. At this preliminary stage, however, single qubit operations gate times
have not been optimised, and we expect single qubit gate times could be reduced
to a point where they are not limiting gate performance. In any case, our
primary objective is to show that two-qubit gate operations in semiconductor
systems can be implemented with times and fidelities comparable to single qubit
operations, and so we have left a more detailed study to future work.

\section{Discussion\label{sec:Conclusions}}

In this paper we have extended the analysis of our proposal for a
two-qubit CPHASE gate between resonant-exchange qubits \cite{Doherty2013},
which uses the large intra-qubit exchange couplings to energetically
suppress leakage caused by inter-qubit exchange couplings. In particular,
we have: demonstrated that high-fidelity two-qubit operations are possible
with gate times comparable to single-qubit operations, provided that gate
times are carefully tuned to account for perturbations beyond first order;
demonstrated that leakage can be further suppressed by using adiabatic
pulse profiles; and shown in conservative simulations that the infidelities
of our two-qubit gate are small (a few percent) even when subject to charge
and Overhauser noise that has been calibrated to recent experiment.

Our two-qubit gate works whenever there is an effective exchange coupling
between the qubits. This could be a direct exchange coupling (as we
have envisaged here), or an indirect coupling through an intermediate
dot which has recently been shown to generate an effective exchange
interaction \cite{Braakman2013,Mehl2014}.

In our analysis, we found (intra-qubit) DC charge noise on the electrodes
defining the quantum dots to be the most potent source of gate infidelity.
While we have chosen our noise parameters to match experimental results,
we do not believe there is any fundamental reason why this noise cannot
be substantially reduced. The next largest source of infidelity was the Overhauser
field. We calibrated our simulations to reflect an unsuppressed Overhauser
field in GaAs/AlGaAs heterostructures. It is possible to suppress
the Overhauser field fluctuations using dynamic nuclear polarisation
\cite{Brataas2014,Neder2014,Nichol2015}, which can reduce the standard
deviation of the field fluctuations by several orders of magnitude
(e.g. \cite{Reilly2008}). It seems reasonable to expect that our
two-qubit gate's fidelities may exceed 99\% even without using echo
pulses. Of course, since low-frequency (DC) noise and the Overhauser
field fluctuations vary very slowly compared to gate times, it should
be reasonably simple to use echo pulses or dynamical decoupling if necessary.

A surprising result, perhaps, is that optimal ratio of inter-qubit
coupling to intra-qubit coupling $J_{c}/J_{z}$ is actually quite
large, especially in the butterfly geometry. As a consequence, it
is not necessary to run these gates particularly slowly, allowing
our gate to run faster (and possibly with greater fidelity) than even
the most carefully constructed pulse sequences (e.g. \cite{Setiawan2014}).

Our approach of energetically suppressing spin-flip transitions in
order to implement two-qubit gates using exchange coupling has utility
in other qubit architectures, as we have already shown in singlet-triplet
qubits \cite{Wardrop2014}. The main benefit of this approach is to
remove dependence on complicated pulse-sequences in order to achieve
high fidelities.

The relative simplicity of our two qubit gate, coupled with its high
performance, commends it for implementation in contemporary experiment.
\begin{acknowledgments}
We acknowledge helpful conversations with Charles Marcus. Research was supported by the Office of the Director of National Intelligence, Intelligence Advanced Research Projects Activity (IARPA), through the Army Research Office grant W911NF-12-1-0354 and by the Australian Research Council (ARC) via the Centre of Excellence in Engineered Quantum Systems (EQuS), project number CE110001013.
\end{acknowledgments}

\bibliographystyle{apsrev4-1}

%

\appendix
\gdef\appendixname{Supplementary}
\onecolumngrid
\newpage

\section{Energy eigenstates for the $m_{z}=1$ subspace}

In this section we list the energy eigenstates for the $m_{z}=1$
subspace of the two resonant-exchange qubit system, in terms of the
single qubit states listed in table \ref{tab:qubit-states} of the
main text. Where an energy degeneracy occurs, we choose states that
maximally exploit the symmetries of the butterfly geometry, in order
to reproduce the splitting and coupling shown in figure \ref{fig:Energy-Levels}
of the main text. In particular, the states of the butterfly configuration
are invariant under global spin rotations, allowing one to write the
degenerate subspace in terms of states with definite angular momentum
using standard Clebsch-Gordan coefficients. The states, along with
their symmetries, are listed in the following table, in eigenstates
are sorted by energy and then by total angular momentum $S$. Note
that energies omit the constant $-B_{\perp}$ contribution shared
by all states. The eigenstate label is chosen to indicate which elements
of the single subspace are involved, and then append a colon followed
by the total angular momentum (with the exception of logical states,
where the total angular momentum is omitted). The ``Parity'' column
indicates the parity accumulated by the state under the operation
involving swapping dot 1 and 3 (the first sign) and 4 and 6 (the second
sign) in the butterfly geometry. $+-$, for example, indicates that
the state is unchanged under swapping the dots in the first qubit,
but attracts a negative sign when swapping the dots of the second.
The ``Swap'' column indicates the parity of the state when interchanging
the qubits, where this is well-defined. Horizontal lines form the
states into groups of equal energy.

\begin{center}
\begin{tabular}{cccccc}
\toprule
Label & State & Energy $(+B_{\perp})$ & $S$ & Parity & Swap\tabularnewline
\midrule
$\ket{QQ:1}$ & $\sqrt{\frac{2}{5}}\ket{Q,Q}-\sqrt{\frac{3}{10}}\left(\ket{Q_{3/2},Q_{-}}+\ket{Q_{-},Q_{3/2}}\right)$ & $0$ & $1$ & $++$ & $+$\tabularnewline
$\ket{QQ:2}$ & $\sqrt{\frac{1}{5}}\left(\ket{Q_{3/2},Q_{-}}-\ket{Q_{-},Q_{3/2}}\right)$ & $0$ & $2$ & $++$ & $-$\tabularnewline
$\ket{QQ:3}$ & $\sqrt{\frac{3}{5}}\ket{Q,Q}+\sqrt{\frac{1}{5}}\left(\ket{Q_{3/2},Q_{-}}+\ket{Q_{-},Q_{3/2}}\right)$ & $0$ & $3$ & $++$ & $+$\tabularnewline
\midrule
$\ket{1Q:1}$ & $\frac{1}{2}\ket{1,Q}-\frac{\sqrt{3}}{2}\ket{1_{-},Q_{3/2}}$ & $-\frac{1}{2}J^{A}_{z}$ & $1$ & $-+$ & \tabularnewline
$\ket{1Q:2}$ & $\frac{\sqrt{3}}{2}\ket{1,Q}+\frac{1}{2}\ket{1_{-},Q_{3/2}}$ & $-\frac{1}{2}J^{A}_{z}$ & $2$ & $-+$ & \tabularnewline
\midrule
$\ket{Q1:1}$ & $\frac{1}{2}\ket{Q,1}-\frac{\sqrt{3}}{2}\ket{Q_{3/2},1_{-}}$ & $-\frac{1}{2}J^{B}_{z}$ & $1$ & $+-$ & \tabularnewline
$\ket{Q1:2}$ & $\frac{\sqrt{3}}{2}\ket{Q,1}+\frac{1}{2}\ket{Q_{3/2},1_{-}}$ & $-\frac{1}{2}J^{B}_{z}$ & $2$ & $+-$ & \tabularnewline
\midrule
$\ket{11}$ & $\ket{1,1}$ & $-\frac{1}{2}(J^{A}_{z}+J^{B}_{z})$ & $1$ & $--$ & $+$\tabularnewline
\midrule
$\ket{0Q:1}$ & $\frac{1}{2}\ket{0,Q}-\frac{\sqrt{3}}{2}\ket{0_{-},Q_{3/2}}$ & $-\frac{3}{2}J^{A}_{z}$ & $1$ & $++$ & \tabularnewline
$\ket{0Q:2}$ & $\frac{\sqrt{3}}{2}\ket{0,Q}+\frac{1}{2}\ket{0_{-},Q_{3/2}}$ & $-\frac{3}{2}J^{A}_{z}$ & $2$ & $++$ & \tabularnewline
\midrule
$\ket{Q0:1}$ & $\frac{1}{2}\ket{Q,0}-\frac{\sqrt{3}}{2}\ket{Q_{3/2},0_{-}}$ & $-\frac{3}{2}J^{B}_{z}$ & $1$ & $++$ & \tabularnewline
$\ket{Q0:2}$ & $\frac{\sqrt{3}}{2}\ket{Q,0}+\frac{1}{2}\ket{Q_{3/2},0_{-}}$ & $-\frac{3}{2}J^{B}_{z}$ & $2$ & $++$ & \tabularnewline
\midrule
$\ket{10}$ & $\ket{1,0}$ & $\frac{1}{2}(J^{A}_{z}+3J^{B}_{z})$ & $1$ & $-+$ & \tabularnewline
\midrule
$\ket{01}$ & $\ket{0,1}$ & $\frac{1}{2}(3J^{A}_{z}+J^{B}_{z})$ & $1$ & $+-$ & \tabularnewline
\midrule
$\ket{00}$ & $\ket{0,0}$ & $\frac{3}{2}(J^{A}_{z}+J^{B}_{z})$ & $1$ & $++$ & $+$\tabularnewline
\bottomrule
\end{tabular}
\par\end{center}

\section{Derivation of the Effective Hamiltonian}

In this section we describe how the effective Hamiltonian of equations
1 and 2 in the main text is lifted from perturbation theory, the procedure
for which follows standard practice \cite{Cohen-Tannoudji2008}. The
main idea is that, in the limit that inter-qubit coupling $J_{c}$
is small compared to intra-qubit coupling $J_{z}$, the evolution
of the logical subspace in which we are interested should be well-approximated
by low-order terms in a perturbation expansion around $J_{c}=0$;
from which considerable insight might be gained into qubit dynamics.

To generate the effective Hamiltonian for a given geometry, we perturb
the Hamiltonian describing the two decoupled resonant exchange qubits
with the Hamiltonian describing the two-qubit coupling of that geometry,
to first order in a Rayleigh-Schroedinger perturbation expansion.
We then construct the effective Hamiltonian using the resulting energies
$E$ and eigenvectors $\ket{E}$ that adiabatically map to the logical
states using $H_{\mathrm{eff}}=\sum_{E}E\ket{E}\bra{E}$. This algorithm
explicitly disregards any leakage operations, but captures the dominant
dynamics on the logical subspace. The result is a diagonal Hamiltonian,
except where the original logical states were degenerate.

For our two resonant exchange qubit system, in which the $\ket{01}$
and $\ket{10}$ states are degenerate (shown in figure 3 of the main
text), this results in a block-diagonal Hamiltonian when written in
the basis $\{\ket{00},\ket{01},\ket{10},\ket{11}\}$. We explicitly
compute the effective Hamiltonians for the geometries shown in figure
2 of the main text; that is, the linear, butterfly and rectangular
geometries respectively. We also explicitly compute the coefficients
of $\sigma_{I}^{A}\sigma_{I}^{B}$, $\sigma_{x}^{A}\sigma_{x}^{B}$,
$\sigma_{y}^{A}\sigma_{y}^{B}$, and $\sigma_{z}^{A}\sigma_{z}^{B}$,
which are then used to populate table II of the main text.

\subsection{Linear}

{\tiny{}
\[
H_{\mathrm{eff}}=\left[\begin{matrix}-\frac{5J_{34}}{36}-\frac{3J^{A}_{z}}{2}-\frac{3J^{B}_{z}}{2} & 0 & 0 & 0\\
0 & -\frac{J_{34}}{4}-\frac{3J^{A}_{z}}{2}-\frac{J^{B}_{z}}{2} & -\frac{J_{34}}{12} & 0\\
0 & -\frac{J_{34}}{12} & -\frac{J_{34}}{4}-\frac{J^{A}_{z}}{2}-\frac{3J^{B}_{z}}{2} & 0\\
0 & 0 & 0 & -\frac{J_{34}}{4}-\frac{J^{A}_{z}}{2}-\frac{J^{B}_{z}}{2}
\end{matrix}\right]
\]
}{\tiny \par}

\[
\mathrm{coefficients}=\left[\begin{matrix}-\frac{2J_{34}}{9}-J^{A}_{z}-J^{B}_{z}\\
\frac{J_{34}}{36}-\frac{J^{B}_{z}}{2}\\
\frac{J_{34}}{36}-\frac{J^{A}_{z}}{2}\\
\frac{J_{34}}{36}
\end{matrix}\right]
\]

\subsection{Butterfly}

{\tiny{}
\[
H_{\mathrm{eff}}=\left[\begin{matrix}-\frac{2J_{25}}{9}-\frac{3J^{A}_{z}}{2}-\frac{3J^{B}_{z}}{2} & 0 & 0 & 0\\
0 & -\frac{J_{25}}{3}-\frac{3J^{A}_{z}}{2}-\frac{J^{B}_{z}}{2} & 0 & 0\\
0 & 0 & -\frac{J_{25}}{3}-\frac{J^{A}_{z}}{2}-\frac{3J^{B}_{z}}{2} & 0\\
0 & 0 & 0 & -\frac{J^{A}_{z}}{2}-\frac{J^{B}_{z}}{2}
\end{matrix}\right]
\]
}{\tiny \par}

\[
\mathrm{coefficients}=\left[\begin{matrix}-\frac{2J_{25}}{9}-J^{A}_{z}-J^{B}_{z}\\
-\frac{J_{25}}{18}-\frac{J^{B}_{z}}{2}\\
-\frac{J_{25}}{18}-\frac{J^{A}_{z}}{2}\\
\frac{J_{25}}{9}
\end{matrix}\right]
\]

\subsection{Rectangular}

{\tiny{}
\[
H_{\mathrm{eff}}=\left[\begin{matrix}-\frac{5J_{16}}{36}-\frac{2J_{25}}{9}-\frac{5J_{34}}{36}-\frac{3J^{A}_{z}}{2}-\frac{3J^{B}_{z}}{2} & 0 & 0 & 0\\
0 & -\frac{J_{16}}{4}-\frac{J_{25}}{3}-\frac{J_{34}}{4}-\frac{3J^{A}_{z}}{2}-\frac{J^{B}_{z}}{2} & -\frac{J_{16}}{12}-\frac{J_{34}}{12} & 0\\
0 & -\frac{J_{16}}{12}-\frac{J_{34}}{12} & -\frac{J_{16}}{4}-\frac{J_{25}}{3}-\frac{J_{34}}{4}-\frac{J^{A}_{z}}{2}-\frac{3J^{B}_{z}}{2} & 0\\
0 & 0 & 0 & -\frac{J_{16}}{4}-\frac{J_{34}}{4}-\frac{J^{A}_{z}}{2}-\frac{J^{B}_{z}}{2}
\end{matrix}\right]
\]
}{\tiny \par}

\[
\mathrm{coefficients}=\left[\begin{matrix}-\frac{2J_{16}}{9}-\frac{2J_{25}}{9}-\frac{2J_{34}}{9}-J^{A}_{z}-J^{B}_{z}\\
\frac{J_{16}}{36}-\frac{J_{25}}{18}+\frac{J_{34}}{36}-\frac{J^{B}_{z}}{2}\\
\frac{J_{16}}{36}-\frac{J_{25}}{18}+\frac{J_{34}}{36}-\frac{J^{A}_{z}}{2}\\
\frac{J_{16}}{36}+\frac{J_{25}}{9}+\frac{J_{34}}{36}
\end{matrix}\right]
\]

Note that to recover the provided coefficients in table II of the
main text, we assume that all inter-qubit couplings ($J_{16}$, $J_{25}$
and $J_{34}$) are equal to $J_{c}$.

\section{Estimating Ideal Gate Time}

\begin{figure}
\includegraphics[width=0.5\columnwidth]{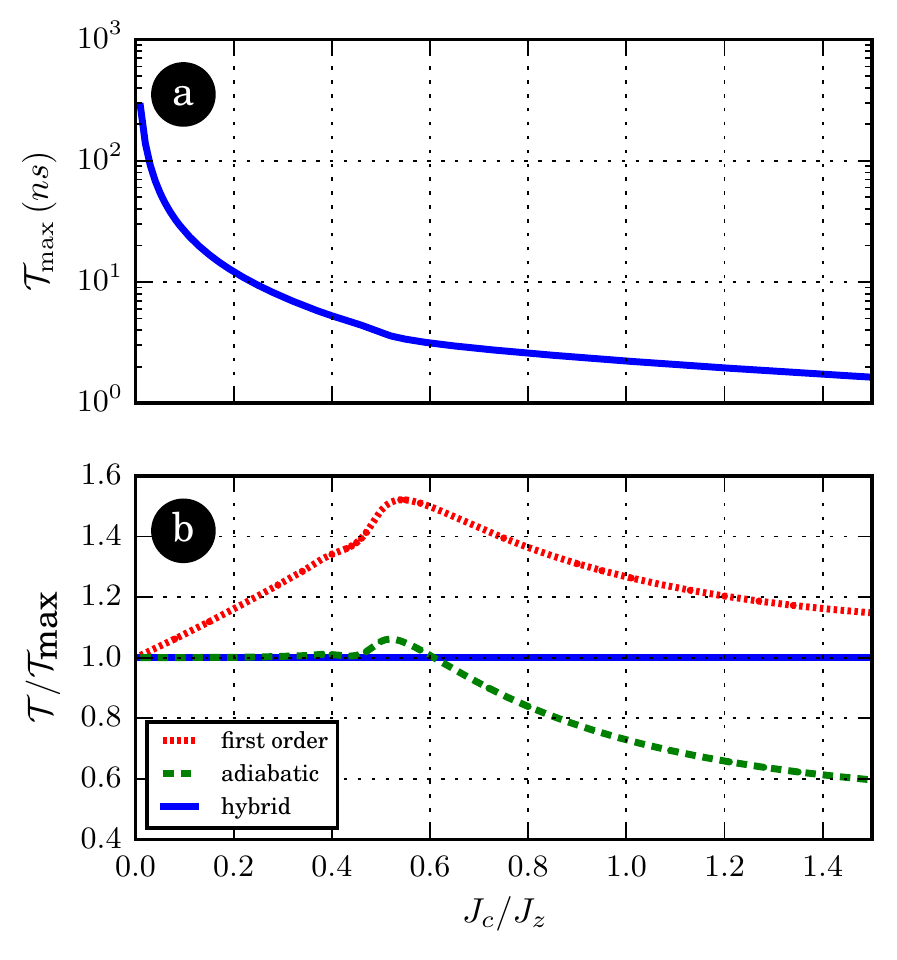}

\protect\caption{(colour online) In (a) we plot the ideal gate time $\mathcal{T}_{\mathrm{max}}$
as computed using the ``hybrid'' method described in the text, which
is our best approach for maximising the fidelity of gate operations.
$\mathcal{T}_{\mathrm{max}}(J_{c}/J_{z})$ is defined as in the text,
and plotted here for $J_{z}^{\mathrm{ref}}=1.65\,\mu eV$. In (b)
we plot the relative difference $\mathcal{T}/\mathcal{T}_{\mathrm{max}}$
of the other methods described in the text, namely ``adiabatic''
(green dashed) and ``first order'' (red dotted) methods, to the
``hybrid'' estimate (blue solid). The plots share the same x axis,
which is the ratio of intra-qubit coupling to inter-qubit coupling
$J_{c}/J_{z}$. Note that the first order approximation consistently
underestimates the phase accrual rate, and so overestimates the gate
time. While an adiabatic phase estimation does better, it ends up
underestimating the phase due to the importance of evolution off the
logical subspace for large enough $J_{c}$.\label{fig:timing-errors}}
\end{figure}

As mentioned in section IVA, over- or under-estimating the ideal gate
time $\tau$ leads to over- or under-accrual of two-qubit phase, resulting
in poor gate fidelities. In particular, using a first order approximation
neglects higher order terms that give rise to convex non-linearity
in $J_{zz}(J_{c})$, and so consistently exaggerates gate times. In
this section, we describe the method by which we obtain accurate timing
estimates for use in the simulations of the main text.

There are two methods that one might employ to find an estimate of
$\tau$ that includes high order perturbations. One can make the assumption
that evolution is perfectly adiabatic, solve for the eigenvalues at
different values of inter-qubit coupling $J_{c}$, and thus infer
the two-qubit phase accumulation rate $J_{zz}$; or, alternatively,
one can locally maximise the gate fidelity over gate times. The first
method works well for small values of $J_{c}/J_{z}$ (where the adiabatic
approximation make sense), and the second method works well for larger
values of $J_{c}/J_{z}$ (where gate times are shorter, and numerical
integration has less time to accumulate error). We therefore use a
hybrid approach that works well across all values of $J_{c}/J_{z}$:
for $J_{c}/J_{z}\le0.15$, we use the adiabatic approach, and for
$J_{c}/J_{z}>0.15$ we use the adiabatic approach to seed a numerical
optimisation of gate fidelity over gate times.

As noted in the main text, the non-linearity of $J_{zz}(J_{c})$ means
that we must repeat this estimation process for each pulse shape considered.
As the dynamics of our (noiseless) two-qubit system is determined
by the ratio of $J_{c}/J_{z}$, given a particular pulse shape, we
need only optimise over the ratio of the two physical degrees of freedom,
rather than both. We therefore proxy the optimisation of $\tau(J_{c},J_{z})$
for any given pulse shape by optimisations over the single parameter
function $\mathcal{T}(J_{c}/J_{z})$. In practice we fix a value of
$J_{z}=J_{z}^{\mathrm{ref}}$, and vary $J_{c}$. $\tau$ can be recovered
from $\mathcal{T}$ using:
\[
\tau(J_{c},J_{z})=\frac{J_{z}^{\mathrm{ref}}}{J_{z}}\mathcal{T}(J_{c}/J_{z}).
\]
The gate time for maximum fidelity $\mathcal{T}_{\mathrm{max}}$ for
the butterfly configuration using a sinusoidal adiabatic pulse profile
is shown in figure \ref{fig:timing-errors}, along with the relative
error of the first order and adiabatic estimations. We note that the
first order approximation consistently underestimates the phase accrual
rate, and so overestimates the gate time. While an adiabatic phase
estimation does better at first, it ends up underestimating the phase
due to the importance of evolution off the logical subspace for large
enough $J_{c}$.

\section{Accelerating Monte-Carlo Convergence in Overhauser Simulations}

Simulations involving the Overhauser field modelled as described in
the main text requires averaging over a multi-variate normal distribution
in 6 variables (11 if DC charge noise is also considered). While a
naive Monte-Carlo simulation that samples the local Overhauser field
contribution for each dot will converge eventually, we speed up the
convergence by instead sampling preferentially from linear combinations
of local Overhauser contributions that appear at lower order in perturbation
theory.

Inspired by perturbation theory, we form a spanning basis for the
magnetic fields:

\begin{eqnarray*}
B_{0} & = & \frac{1}{6}\sum_{n}B_{n}\\
\Delta B & = & \frac{1}{3}\left(\sum_{n\le3}B_{n}-\sum_{n\ge4}B_{n}\right)\\
\Delta_{ij} & = & B_{i}-B_{j}\,\text{for \ensuremath{\Delta_{13}} and \ensuremath{\Delta_{46}}}\\
\Delta_{ijk} & = & B_{i}-2B_{j}+B_{k}\,\text{for \ensuremath{\Delta_{123}} and \ensuremath{\Delta_{456}}}
\end{eqnarray*}
where $B_{n}$ (with $n\in[1,6]$) is the z-component of the Overhauser
field at dot $n$. We learn from perturbation theory that $\Delta_{123}$
and $\Delta_{456}$ first contribute at zeroth order in perturbation
theory, $\Delta_{13}$ and $\Delta_{46}$ at first order, and $\Delta B$
at fourth order. As $B_{0}$ is a global field, it does not contribute.

The higher the order at which the terms contribute, the more significant
that term is to the dynamics of the qubit system. It makes sense,
therefore, to sample lower order terms more often. Technically, one
should weight each term roughly as $(J_{c}/J_{z})^{o}$, where $o$
is the order at which the term appears in perturbation theory. We
found it simpler, however, to use a conservative ratio of $5$ between
terms of different order. That is, for each 100 simulations, we sample
$\Delta_{123}$ and $\Delta_{456}$ 100 times, $\Delta_{13}$ and
$\Delta_{46}$ 20 times, and $\Delta B$ 4 times.

As the terms of the new basis are linear compositions of the $B_{n}$
terms, the random variable associated with each term will differ from
the underlying local field fluctuations. In particular, if each $B_{n}\sim\mathcal{N}(B_{0},\sigma_{B}^{2})$,
then the relevant terms of new basis will be sampled from the distributions:
\begin{eqnarray*}
\Delta_{ijk} & \sim & \mathcal{N}(0,4\sigma_{B}^{2})\\
\Delta_{ij} & \sim & \mathcal{N}(0,2\sigma_{B}^{2})\\
\Delta B & \sim & \mathcal{N}(0,2\sigma_{B}^{2})
\end{eqnarray*}

\section{Cross-Coupling}

Perturbation theory provides the insight that beyond first order the
Overhauser field cross-couples with charge noise in $J_{c}$ at second
order and above. This allows the Overhauser field to contribute non-trivially
to two-qubit phases. For example, the two-qubit phase accumulation
rate to second order for the butterfly configuration is given by:
\begin{eqnarray*}
J_{zz} & = & \frac{1}{9}J_{c}+J_{c}^{2}\frac{8(J^{A}_{z}+J^{B}_{z})^{2}-5J^{A}_{z}J^{B}_{z}}{243J^{A}_{z}J^{B}_{z}\left(J^{A}_{z}+J^{B}_{z}\right)}\\
{\color{red}\star} &  & {\color{red}-\frac{2}{81}J_{c}\left(\frac{\Delta_{123}}{J^{A}_{z}}+\frac{\Delta_{456}}{J^{B}_{z}}\right)}\\
 &  & +\mathcal{O}(J_{c}\Delta^{2})+\mathcal{O}(J_{c}^{2}\Delta)+\mathcal{O}(J_{c}^{3}),
\end{eqnarray*}
where $\Delta_{ijk}=B_{i}-2B_{j}+B_{k}$, and where the coloured and
starred line is the effect of cross-coupling. The intuition is that
as $J_{c}/J_{z}$ and the $\sigma_{B}$ increase, the greater the
two-qubit cross-coupling error. For the useful range of $J_{c}/J_{z}$
values, and the experimentally inspired values of $\sigma_{B}$ used
in the main text, the effect of this cross-coupling is negligible.
\end{document}